\documentclass[romanindices,journal]{IEEEtran}
\usepackage[english]{babel}
\usepackage{amssymb}
\usepackage[cmex10]{amsmath}
\usepackage{overpic}
\usepackage{algorithm}
\usepackage{algorithmic}
\usepackage{multirow}
\usepackage{array}
\usepackage{graphicx}
\usepackage{bm}
\usepackage{color}
\usepackage{tikz}
\usepackage{epstopdf}
\usepackage{stfloats}
\usepackage{cite}
\usepackage{tikz}
\usepackage{amsmath}
\usepackage{psfrag}
\usepackage{acronym}
\usepackage[caption=false,font=scriptsize]{subfig}
\usepackage{enumerate}
\usetikzlibrary{arrows}
\usetikzlibrary{decorations}

\definecolor{BLUE}{rgb}{0,0,1}

\newtheorem{proposition}{Proposition}
\newtheorem{remark}{Remark}

\acrodef{aoa}[AOA]{angle-of-arrival}
\acrodef{acf}[ACF]{autocorrelation function}
\acrodef{bcrb}[BCRB]{Bayesian Cram\'{e}r-Rao bound}
\acrodef{bp}[BP]{belief propagation}
\acrodef{cdi}[CDI]{cooperative dilution intensity}
\acrodef{cl}[CL]{cooperative localization}
\acrodef{cdf}[CDF]{cumulative distribution function}
\acrodef{crb}[CRB]{Cram\'{e}r-Rao bound}
\acrodef{crlb}[CRLB]{Cram\'{e}r-Rao lower bound}
\acrodef{dof}[DoF]{degree of freedom}
\acrodef{dct}[DCT]{discrete cosine transform}
\acrodef{dpeb}[DPEB]{directional position error bound}
\acrodef{fim}[FIM]{Fisher information matrix}
\acrodef{efim}[EFIM]{equivalent Fisher information matrix}
\acrodef{ici}[ICI]{information coupling intensity}
\acrodef{icrb}[ICRB]{inverse CRB}
\acrodef{iid}[i.i.d.]{independently and identically distributed}
\acrodef{isac}[ISAC]{Integrated Sensing and Communication}
\acrodef{mse}[MSE]{mean-squared error}
\acrodef{pdf}[PDF]{probability density function}
\acrodef{peb}[PEB]{position error bound}
\acrodef{speb}[SPEB]{squared position error bound}
\acrodef{pll}[PLL]{phase-locked loop}
\acrodef{rbs}[RBS]{reference broadcast synchronization}
\acrodef{rhs}[RHS]{right hand side}
\acrodef{rii}[RII]{ranging information intensity}
\acrodef{rss}[RSS]{received signal strength}
\acrodef{rc}[RC]{ranging coefficient}
\acrodef{speb}[SPEB]{squared position error bound}
\acrodef{toa}[TOA]{time-of-arrival}
\acrodef{tdoa}[TDOA]{time-difference-of-arrival}
\acrodef{tpsn}[TPSN]{time synchronization protocol for sensor network}
\acrodef{vmp}[VMP]{variational message passing}
\acrodef{wsn}[WSN]{wireless sensor network}
\acrodef{efim}[EFIM]{equivalent Fisher information matrix}
\acrodef{dio}[DIO]{distance-information-only}
\acrodef{aio}[AIO]{angle-information-only}
\acrodef{saaf}[SAAF]{squared array aperture function}
\acrodef{snc}[S\&C]{sensing and communications}
\acrodef{uoa}[UOA]{uniformly oriented array}
\acrodef{rgg}[RGG]{random geometric graph}
\acrodef{rms}[RMS]{root-mean-square}
\acrodef{snr}[SNR]{signal-to-noise ratio}
\acrodef{eoc}[EoC]{efficiency of cooperation}
\acrodef{npi}[NPI]{nominal position information}
\acrodef{gnss}[GNSS]{global navigation satellite system}
\acrodef{mimo}[MIMO]{multiple-input multiple-output}
\acrodef{mcs}[MCS]{minimally constrained system}
\acrodef{zzb}[ZZB]{Ziv-Zakai bound}
\acrodef{wwb}[WWB]{Weiss-Weinstein lower bound}
\acrodef{nlos}[NLOS]{non-light-of-sight}
\acrodef{mmse}[MMSE]{minimum mean squared error}
\acrodef{uav}[UAV]{unmanned aerial vehicle}
\acrodef{ppp}[PPP]{Poisson point process}
\acrodef{bpp}[BPP]{binomial point process}
\acrodef{cln}[CLN]{cooperative location-aware network}
\acrodef{pdr}[PDR]{pedestrian dead reckoning}
\acrodef{ml}[ML]{maximum likelihood}
\acrodef{map}[MAP]{maximum \textit{a posteriori}}

\acrodefplural{dof}[DoFs]{degrees of freedom}

\usepackage{winsnotation}
\usepackage{flushend}

\title{SNR-Adaptive Ranging Waveform Design Based on \\Ziv-Zakai Bound Optimization}
\author{Yifeng Xiong, \IEEEmembership{Member, IEEE}, and Fan Liu, \IEEEmembership{Member, IEEE}
\thanks{Y. Xiong is with the School of Information and Communication Engineering, Beijing University of Posts and Telecommunications, Beijing 100876, China. (e-mail: yifengxiong@bupt.edu.cn)}
\thanks{F. Liu is with the Department of Electronic and Electrical Engineering, Southern University of Science and Technology, Shenzhen 518055, China. (e-mail: liuf6@sustech.edu.cn)}
}

\begin{document}
\maketitle

\begin{abstract}
Location-awareness is essential in various wireless applications. The capability of performing precise ranging is substantial in achieving high-accuracy localization. Due to the notorious ambiguity phenomenon, optimal ranging waveforms should be adaptive to the signal-to-noise ratio (SNR). In this letter, we propose to use the Ziv-Zakai bound (ZZB) as the ranging performance metric, as well as an associated waveform design algorithm having theoretical guarantee of achieving the optimal ZZB at a given SNR. Numerical results suggest that, in stark contrast to the well-known high-SNR design philosophy, the detection probability of the ranging signal becomes more important than the resolution in the low-SNR regime.
\end{abstract}


\section{Introduction}
\IEEEPARstart{L}{ocation} information acquisition is one of the most indispensable sensing capabilities in wireless networks, which serves as a key enabler of a wide range of tasks including autonomous driving, internet of things, logistic and health care \cite{proc_win,cooploc,av,iot1}. Precise ranging is at the core of high-accuracy localization systems, for which waveform design is an essential task. Regarding ranging waveform design, a key observation is that the time-resolution of a ranging signal is roughly proportional to the reciprocal of its bandwidth, which has inspired many existing ranging techniques, including techniques based on ultra-wideband (UWB) signals \cite{poor_msp,fundamental1}, as well as those benefitted from carrier aggregation \cite{ca_ranging,ca_ranging2}. This insight has also given rise to considerable efforts devoted to network operation techniques yielding optimized bandwidth allocation strategies for localization systems \cite{netloc_win,twc_bandwidth,tcom_bandwidth}.

\begin{figure}[t]
\centering
\includegraphics[width=.45\textwidth]{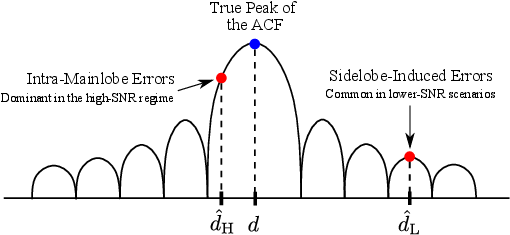}
\caption{A sketchy illustration of the signal \ac{acf} and different types of ranging errors.}
\label{fig:ranging_error}
\end{figure}

From a theoretical perspective, the \ac{crb} of distance corroborates the ``large bandwidth'' intuition, stating that the lowest achievable \ac{mse} is inversely proportional to the squared \ac{rms} bandwidth \cite{fundamental1}. In more precise terms, the \ac{rms} bandwidth is known to be proportional to the curvature of the signal \ac{acf} at the peak \cite{fundamental1}. As it can be observed from Fig.~\ref{fig:ranging_error}, when the \ac{snr} is high, the estimated distance (represented by $\hat{d}_{\rm H}$ in Fig.~\ref{fig:ranging_error}) would lie within the mainlobe around the true peak, and hence the ranging error is mainly determined by the aforementioned curvature. However, when the \ac{snr} is relatively low, there is a non-negligible probability that the maximum matched filter response locates at one of the \ac{acf} sidelobes, resulting in larger ranging errors corresponding to $\hat{d}_{\rm L}$ in Fig.~\ref{fig:ranging_error}. This issue is known as the ``ambiguity phenomenon'' \cite{ambiguity_book}, which suggests that the optimal ranging waveform should be \ac{snr}-adaptive.

Designing \ac{snr}-adaptive ranging waveforms would certainly rely on \ac{snr}-dependent performance metrics. \ac{crb}, being the most widely employed metric for estimators, does depend on \ac{snr}. Unfortunately, \ac{crb} is related to the ranging waveform only via its \ac{rms} bandwidth, suggesting that it does not account for the ambiguity phenomenon. Therefore, \ac{crb}-based waveform design yields a single waveform for any \ac{snr}. In fact, the actual \ac{mse} of the \ac{crb}-optimal waveform never attains the corresponding \ac{crb} (at an arbitrarily high \ac{snr}), as will be detailed later in this letter.

A promising candidate metric accounting for the ambiguity phenomenon is the \ac{zzb} \cite{zzb}. The main underlying idea of \ac{zzb} is that the \ac{mse} can be expressed in terms of the error probability of binary detection problems \cite{ezzb}. Intuitively, in the ranging problem, such binary detection problems concern distinguishing the true \ac{acf} peak from the sidelobes, and hence automatically take the ambiguity phenomenon into account. The \ac{zzb} of the ranging problem has been derived and analyzed extensively in the existing literature \cite{zzb,ezzb,statistical_TOA}, showing that it does reflect the ambiguity issue in practice. However, to the best knowledge of the authors, \ac{zzb} has not been applied to waveform design, mainly due to its lack of closed-form expressions.

In this letter, instead of attempting to find suitable closed-form approximations, we aim for designing \ac{snr}-adaptive ranging waveforms by directly solving the \ac{zzb} optimization problem as a variational problem. Our contributions are summarized as follows.
\begin{itemize}
\item We show that the variational problem of \ac{zzb} optimization for ranging waveform design is a convex problem;
\item We propose a discretized numerical reformulation of the \ac{zzb} optimization problem preserving the convexity, as well as an associated computation-efficient design algorithm based on the gradient projection method;
\item Using numerical examples, we demonstrate that smaller \ac{rms} bandwidth can be more favorable under strong noise, suggesting that detection probability is preferred to resolution in the low-\ac{snr} regime.
\end{itemize}

\textit{Notations}: Throughout the letter, we use $[\V{x}]_i$ to represent the $i$-th entry of vector $\V{x}$, and use $[\M{A}]_{i,j}$ to represent the $(i,j)$-th entry of matrix $\M{A}$.

\section{System Model}
Let us consider the generic time-based ranging model of
\begin{equation}
y(t) = s(t-\tau) + n(t),
\end{equation}
where $y(t)$ denotes the received signal, $s(t)$ represents the transmitted signal, $n(t)$ denotes the zero-mean white Gaussian noise having a constant power spectral density of $N_0$. The term $\tau = d/c$ denotes the propagation delay, with $c$ being the propagation speed (e.g. speed of light), and $d$ being the distance to be estimated. With this model in hand, we may obtain the following likelihood function
\begin{align}
f(y;\tau) &\propto e^{-\frac{1}{N_{\rm 0}}\int_{-\infty}^{\infty} (y(t)-s(t-\tau))^2{\rm d}t}\nonumber \\
&\propto e^{-\frac{2}{N_{\rm 0}} \left(R(\tau)+\int_{-\infty}^{\infty} n(t)s(t-\tau){\rm d}t\right)},
\end{align}
where $R(\tau)$ is the \ac{acf} of $s(t)$ given by $R(\tau) = \int_{-\infty}^{\infty} s(t-\tau)s(t){\rm d}t$.

According to \cite{statistical_TOA}, when the \textit{a priori} distribution of the delay $\tau$ is a uniform distribution over $[\tau_{\min},\tau_{\max}]$, the \ac{zzb} of the distance $d$ takes the following form
\begin{align}\label{zzb}
\mathbb{E}\left\{(d\!-\!\hat{d})^2\right\}&\!\geq\! \zeta_{\rm SNR}[\widetilde{R}] \nonumber \\
:&\!=\! \int_0^{\epsilon_{\max}} x Q\Big(\sqrt{2^{-1}{\rm SNR}(1\!-\!\widetilde{R}(x))}\Big){\rm d}x,
\end{align}
where $\epsilon_{\max}$ is the maximum possible ranging error, namely $\epsilon_{\max}=c(\tau_{\max}-\tau_{\min})/2$, $Q(\cdot)$ denotes the function of $Q(z) = \int_z^{\infty} \frac{1}{\sqrt{2\pi}} e^{-\frac{1}{2}x^2} {\rm d}x$, $\widetilde{R}(x)$ denotes the normalized (and rescaled) \ac{acf} defined as
\begin{equation}
\widetilde{R}(x) = R(x/c)/R(0),
\end{equation}
while ${\rm SNR}$ is the \ac{snr} given by ${\rm SNR} = \frac{1}{N_0} \int_{-\infty}^{\infty} s^2(t) {\rm d}t$.\footnote{A noteworthy fact is that the \ac{zzb} is related to the ranging waveform only through its \ac{acf}. The remaining degrees of freedom may be used to convey information. This characteristic is particularly useful in the context of integrated sensing and communication \cite{tit_isac}.}

\begin{remark}
We are now in the position to illustrate a key difference between \ac{crb}-based waveform design and the proposed \ac{zzb}-based design. Specifically, the \ac{crb} of $d$ is given by \cite{survey_limits}
\begin{equation}\label{crb_ranging}
\mathbb{E}\{(d-\hat{d})^2\}\geq c^2(8\pi^2\beta^2{\rm SNR})^{-1},
\end{equation}
where $\beta=\int_{-\infty}^{\infty}f^2|S(f)|^2{\rm d}f/\int_{-\infty}^{\infty}|S(f)|^2{\rm d}f$ is the \ac{rms} bandwidth, with $S(f)$ being the Fourier transform of $s(t)$. In this expression, the term ${\rm SNR}^{-1}$ is merely a linear proportional factor. Consequently, the \ac{crb}-based optimal waveform does not vary with \ac{snr}. By contrast, observe that in the expression \eqref{zzb} of \ac{zzb}, ${\rm SNR}^{-1}$ no longer affects the \ac{zzb} in a simple linear manner, given the highly non-linear $Q$ function. Therefore, for different \ac{snr} values, the optimal signal ACF $\widetilde{R}(x)$ will be different. This enables us to design \ac{snr}-adaptive normalized \acp{acf}.
\end{remark}

\section{Problem Formulation and the Proposed Design Method}
We consider the problem of \ac{snr}-adaptive waveform design under bandwidth constraint. In this letter, we consider the baseband ranging scenario, and assume that the fractional bandwidth is small.\footnote{The fractional bandwidth is defined as $B_{\rm frac}=B/f_{\rm c}$, where $B$ is the signal bandwidth while $f_{\rm c}$ denotes the carrier frequency. Baseband signal processing is known to be sufficient in the sense of Le Cam's distance when the fractional bandwidth is small \cite[Remark 4]{array_limits}.} For a given design \ac{snr} ${\rm SNR}_{\rm D}$, this problem can be expressed as
\begin{subequations}\label{zzb_original}
\begin{align}
\min_{\widetilde{R}(x)} &~~\zeta_{{\rm SNR}_{\rm D}}[\widetilde{R}]\\
{\rm s.t.}&~~\widetilde{R}(0)=1,~\widetilde{R}(x)\leq 1,~\forall x\in[0,\epsilon_{\max}].\\
&~~\int_{-\infty}^{\infty} \widetilde{R}(x) e^{-j\omega x} {\rm d}x \geq 0,~\forall \omega \in (-\infty,\infty), \label{pdacf}\\
&~~\int_{-\infty}^{\infty} \widetilde{R}(x) e^{-j\omega x} {\rm d}x = 0,~\forall |\omega|> B, \label{bandwidth_constraint}
\end{align}
\end{subequations}
where \eqref{pdacf} follows from the positive semi-definiteness of \acp{acf}, \eqref{bandwidth_constraint} is the bandwidth constraint.\footnote{The bandwidth constraint is a low-pass constraint since we consider baseband waveforms. This is an appropriate treatment when the bandwidth of the baseband signal is much smaller compared to the carrier frequency.} In its original form, \eqref{zzb_original} is a variational problem which does not admit any analytical solution. Nevertheless, we have the following result:
\begin{proposition}[Convexity of Variational \ac{zzb} Optimization]\label{prop:convex_variational}
The objective functional $\zeta_{\rm SNR_D}[\widetilde{R}]$ is a convex functional with respect to $\widetilde{R}(x)$.
\begin{IEEEproof}
Please refer to Appendix \ref{sec:proof_convex_variational}.
\end{IEEEproof}
\end{proposition}

Proposition \ref{prop:convex_variational} inspires us to find convex-preserving numerical reformulations of \eqref{zzb_original}, which guarantee the achievability of the global optimum. In this letter, we consider the specific dicretized formulation as follows:
\begin{subequations}\label{zzb_discretize}
\begin{align}
\min_{\V{r}} &~~\zeta_{{\rm SNR}_{\rm D}}(\V{r}) := \sum_{i=1}^N x_iQ\left(\sqrt{2^{-1}{\rm SNR}_{\rm D}(1\!-\!r_i)}\right)\\
{\rm s.t.}&~~r_1=1, ~\V{r}\preceq \V{1},~\M{C}\V{r}\succeq \V{0}, \label{pdacf_dct}\\
&~~[\M{C}\V{r}]_i = 0,~\forall i> B_{\rm dis}, \label{bandwidth_constraint_dct}
\end{align}
\end{subequations}
where \eqref{bandwidth_constraint_dct} is the discretized version of the bandwidth constraint, $N$ is the number of discretized sample points, $x_i=(i-1)\epsilon_{\max}/(N-1)$ denotes the $i$-th sample point, $\V{r}=[r_1,\dotsc,r_N]^{\rm T}$ is the sampled version of $\widetilde{R}(x)$, and $\M{C}$ is the matrix of \ac{dct} satisfying
\begin{equation}\label{dct}
[\M{C}]_{k,n} = \sqrt{\frac{2}{N}} \cos \left(\frac{\pi}{4N}(2k-1)(2n-1)\right).
\end{equation}

Next, we show that this discretization is indeed convex-preserving.
\begin{proposition}\label{prop:convex}
The problem \eqref{zzb_discretize} is convex.
\begin{IEEEproof}
(Sketch) Computing the Hessian matrix of $\zeta_{{\rm SNR}_{\rm D}}(\V{r})$ with respect to $\V{r}$, and by noticing that $\frac{\partial^2 \zeta_{{\rm SNR}_{\rm D}}(\V{r})}{\partial r_i \partial r_j}=0$ holds for any $i\neq j$, we see that the Hessian matrix is positive semi-definite, and hence $\zeta_{{\rm SNR}_{\rm D}}(\V{r})$ is convex. This further implies that the problem \eqref{zzb_discretize} is convex, since the feasible set characterized by \eqref{pdacf_dct} and \eqref{bandwidth_constraint_dct} is a convex polytope.
\end{IEEEproof}
\end{proposition}

In light of Proposition \ref{prop:convex}, we can obtain the global optimal solution of \eqref{zzb_discretize} using the gradient projection method. The projection is practically implementable since the Hessian matrix is diagonal and can be readily calculated using \eqref{second_variation}. Specifically, the algorithm iterates as follows
\begin{subequations}
\begin{align}
\V{u}^{(l)} &= \Set{P}_{\Set{S}}\big(\V{r}^{(l-1)}-\sigma\nabla \zeta_{{\rm SNR}_{\rm D}}(\V{r}^{(l-1)})\big), \\
\V{r}^{(l)} &= \V{r}^{(l-1)} + \alpha_{l-1}(\V{u}^{(l)} - \V{r}^{(l-1)}),
 \end{align}
\end{subequations}
until convergence, where $\sigma$ is a fixed step size, while $\alpha_{l-1}$ is a variable step size satisfying the Armijo's condition \cite{armijo}. The projection operator $\Set{P}_{\Set{S}}(\V{x})$ is defined by the following optimization problem
\begin{equation}\label{projection}
\Set{P}_{\Set{F}}(\V{x}) = \mathop{\rm argmin}_{\V{r}}~\|\V{x}-\V{r}\|^2,~{\rm s.t.}~\V{r}\in\Set{S},
\end{equation}
where the feasible set $\Set{S}$ is given by
$$
\Set{S} = \left\{\V{r}|r_1=1,\V{r}\preceq \V{1},\M{C}\V{r}\succeq\V{0},[\M{C}\V{r}]_i=0,\forall i>B_{\rm dis}\right\}.
$$

Although problem \eqref{projection} can be solved by general-purpose convex optimization solvers (e.g. CVX \cite{cvx}), such solvers may be computationally inefficient. To this end, we apply Dykstra's projection method \cite{dykstra} to simplify the procedure. To elaborate, we compute $\Set{P}_{\Set{S}}(\V{x})$ by iteratively applying the following rules
\begin{subequations}
\begin{align}
\V{a}_k &= \Set{P}_{\Set{T}}(\V{x}_k+\V{p}_k),~\V{p}_{k+1}=\V{x}_k+\V{p}_k-\V{a}_k,\\
\V{x}_{k+1}&=\Set{P}_{\Set{F}}(\V{a}_k+\V{q}_k),~\V{q}_{k+1}=\V{a}_k+\V{q}_k-\V{x}_{k+1},
\end{align}
\end{subequations}
ensuring that $\V{x}_k$ converges to $\Set{P}_{\Set{S}}(\V{x})$ as $k$ increases, where the notations $\V{p}_k$ and $\V{q}_k$ are temporary intermediate variables, initialized by $\V{p}_0=\V{q}_0=\V{0}$, $\V{x}_0=\V{x}$, the ``temporal domain feasible set'' $\Set{T}$ and the ``frequency domain feasible set'' $\Set{F}$ are defined as
\begin{subequations}
\begin{align}
\Set{T} &= \left\{\V{r} | r_1=1,\V{r}\preceq \V{1}\right\},\\
\Set{F} &= \left\{\V{r}|\M{C}\V{r}\succeq \V{0},[\M{C}\V{r}]_i=0,\forall i>B_{\rm dis}\right\},
\end{align}
\end{subequations}
respectively. The corresponding projection operators are characterized by
\begin{subequations}
\begin{align}
{[\Set{P}_{\Set{T}}(\V{x})]}_i &= \left\{
                                  \begin{array}{ll}
                                    1, & \hbox{$i=1$ or $i\neq 1,~r_i>1$;} \\
                                    x_i, & \hbox{otherwise,}
                                  \end{array}
                                \right.\\
{[\M{C}\Set{P}_{\Set{F}}(\V{x})]}_i &= \left\{
                                       \begin{array}{ll}
                                         0, & \hbox{$i>B_{\rm dis}$ or $[\M{C}\V{x}]_i<0$;} \\
                                         {[\M{C}\V{x}]}_i, & \hbox{otherwise.}
                                       \end{array}
                                     \right.
\end{align}
\end{subequations}
Overall, the algorithm requires only $\rm SNR_D$, the fixed step size $\sigma$ and an initial guess $\V{r}^{(0)}$ as inputs. The variable step sizes $\{\alpha_l\}$ can be determined at runtime. The workflow of the algorithm is summarized in Alg. \ref{alg:zzb}.
\begin{remark}
Denoting the number of Dykstra's iteration as $K$ (as used in Alg. \ref{alg:zzb}), the complexity of our projection method is on the order of $O(KN\log N)$, by relying on fast \ac{dct}. By contrast, typical off-the-shelf solvers based on interior-point method would yield a complexity of $O(N^{3.5})$. This elucidates the computational efficiency of the proposed method.
\end{remark}

\begin{algorithm}[t]
    \caption{The proposed waveform design algorithm.}
    \begin{algorithmic}[1]\label{alg:zzb}
        \STATE Initialize $\V{r}^{(0)}$;
        \FOR{$l=1$:$L$}
        \STATE Compute $\V{x}_0=\V{r}^{(l-1)}-\sigma \nabla \zeta_{{\rm SNR}_{\rm D}}(\V{r}^{(l-1)})$;
        \STATE Initialize $\V{p}_0=\V{q}_0=\V{0}$;
        \FOR{$k=1$:$K$}
        \STATE $\V{a}_k = \Set{P}_{\Set{T}}(\V{x}_k+\V{p}_k),~\V{p}_{k+1}=\V{x}_k+\V{p}_k-\V{a}_k$;
        \STATE $\V{x}_{k+1}=\Set{P}_{\Set{F}}(\V{a}_k+\V{q}_k),~\V{q}_{k+1}=\V{a}_k+\V{q}_k-\V{x}_{k+1}$;
        \ENDFOR;
        \STATE Let $\V{u}^{(l)}=\V{x}_K$;
        \STATE Compute $\V{r}^{(l)}=\V{r}^{(l-1)}+\alpha_{l-1}\left(\V{u}^{(l)}-\V{r}^{(l-1)}\right)$;
        \ENDFOR;
    \end{algorithmic}
\end{algorithm}

\section{Numerical Results}\label{sec:numerical}
Throughout the numerical examples, we consider the scenario where $N=1000$,\footnote{$N=1000$ is chosen such that the discretization error is negligible.} $B_{\rm dis}=40$. The initial guess $\V{r}^{(0)}$ is set to be the Sinc pulse $\V{s}$ satisfying
\begin{equation}
[\M{C}\V{s}]_i=\left\{
               \begin{array}{ll}
                 c_{\rm s}, & \hbox{$i\leq B_{\rm dis}$;} \\
                 0, & \hbox{$i>B_{\rm dis}$,}
               \end{array}
             \right.
\end{equation}
where $c_{\rm s}$ is a constant ensuring that $s_i=1$. We also use the Sinc pulse as a performance benchmark since it corresponds to a uniform frequency-domain power allocation strategy. We assume that both the transmitter and the receiver locate within the interval $[-1,1]$ on the real line, and thus the maximum distance estimation error is $\epsilon_{\max}=2$.\footnote{The unit of the distance is neglected as it is unnecessary for our purpose. We also use the convention of $c=1$ for numerical convenience.}

\begin{figure}[t]
\subfloat[][\ac{zzb}-optimal waveforms at different design \ac{snr} ${\rm SNR}_{\rm D}$.]{
\centering
\includegraphics[width=.9\columnwidth]{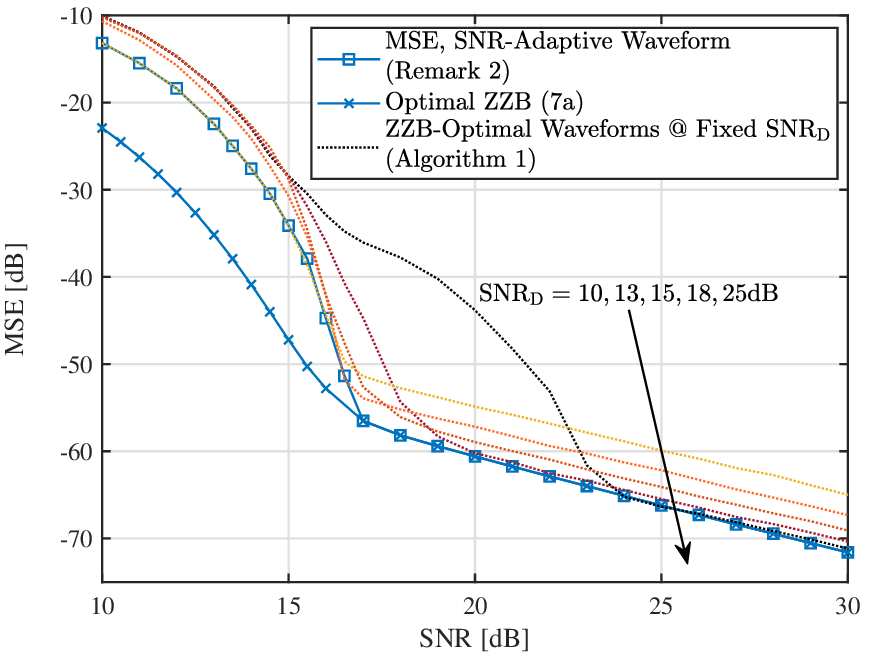}
\label{fig:zzb_at_snrd}
}\\
\subfloat[][\ac{snr}-adaptive waveforms vs. Sinc pulse, \ac{crb} and \ac{zzb}.]{
\centering
\includegraphics[width=.9\columnwidth]{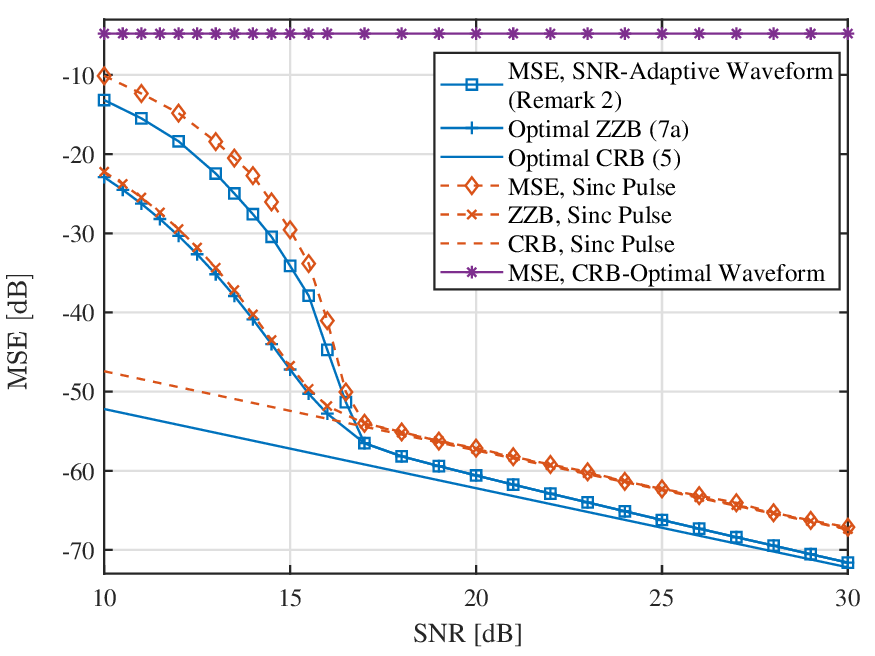}
\label{fig:zzb_vs_mmse_40}
}
\caption{The \ac{mse} of the designed waveforms and benchmarks.}
\end{figure}

Before delving into details, we note that although the proposed algorithm designs the waveform that achieves the lowest \ac{zzb} at a given design \ac{snr} $\rm SNR_D$, the \ac{zzb}-optimal waveform at a specific $\rm SNR_D$ does not necessarily achieve the optimal \ac{mse} at that \ac{snr}, since \ac{zzb} is a only lower bound of the \ac{mse}. In light of this, we see that
\begin{remark}[Adaptive Strategy]
We may conceive \ac{snr}-adaptive waveforms by selecting the \ac{zzb}-optimal waveform achieving the lowest \ac{mse} at each $\rm SNR_D$.
\end{remark}

We first demonstrate the optimal \ac{zzb} achieved by the proposed method, portrayed in Fig.~\ref{fig:zzb_at_snrd}. The first dotted line intersecting with the arrow in Fig.~\ref{fig:zzb_at_snrd} represent ${\rm SNR}_{\rm D}=10$dB, the second denotes ${\rm SNR}_{\rm D}=13$dB, and so forth. The \ac{mse} achieved by certain \ac{zzb}-optimal waveforms at fixed $\rm SNR_D$, and that achieved using the adaptive strategy are also incorporated. Observe that the \ac{zzb} is achievable at relatively high \ac{snr} (i.e. larger than around $16$dB). Another noteworthy issue is that the \ac{zzb}-optimal waveform is indeed not necessarily \ac{mse}-optimal at ${\rm SNR}_{\rm D}$. For example, in Fig.~\ref{fig:zzb_at_snrd}, the \ac{zzb}-optimal waveform at ${\rm SNR_D}=18$dB is not \ac{mse}-optimal until ${\rm SNR_D}=20$dB, while the \ac{zzb}-optimal waveform at ${\rm SNR_D}=13$dB is \ac{mse}-optimal around ${\rm SNR_D}=16.5$dB.

\begin{figure}[t]
\subfloat[][Power spectral densities.]{
\centering
\includegraphics[width=.9\columnwidth]{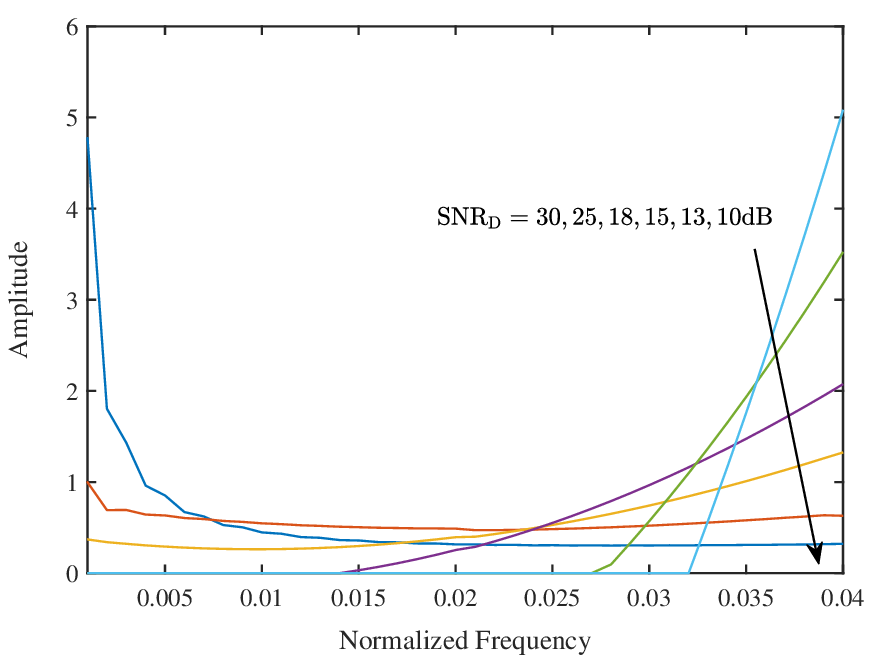}
\label{fig:zzb_optimal_psd}
}\\
\subfloat[][Normalized \acp{acf}.]{
\centering
\includegraphics[width=.9\columnwidth]{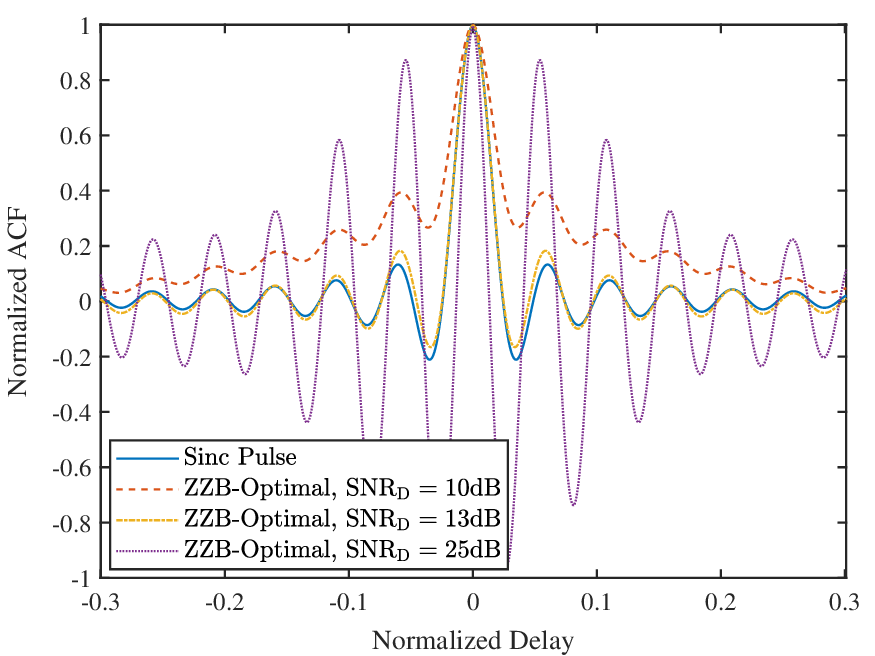}
\label{fig:zzb_optimal_acf}
}
\caption{The frequency- and time-domain representations of the \ac{zzb}-optimal waveforms at various design \acp{snr}.}
\end{figure}

Next, let us compare the performance of the proposed \ac{snr}-adaptive waveforms with that of the Sinc pulse, as portrayed in Fig.~\ref{fig:zzb_vs_mmse_40}. Observe that the Sinc pulse achieves its corresponding \ac{crb} at relatively high \acp{snr}. By contrast, the optimal \ac{zzb} does not attain the optimal \ac{crb} even at ${\rm SNR}=30$dB, but the gap between them shrinks as the \ac{snr} increases. Another closely-related phenomenon is that the \ac{mse} of the \ac{crb}-optimal waveform is unfavorable over the entire \ac{snr} range. To understand this phenomenon, first note that the \ac{crb}-optimal waveform $\V{r}_{\rm CRB}$ is a sinusoidal signal, namely $[\M{C}\V{r}_{\rm CRB}]_i=1$ for $i=B_{\rm dis}$ and $[\M{C}\V{r}_{\rm CRB}]_i=0$ otherwise, according to \eqref{crb_ranging}. This waveform will never exhibit a \ac{crb}-achieving \ac{mse} at any \ac{snr}, since it is periodical and hence there will be unresolvable integer ambiguities resulting in uniformly distributed ranging errors across the entire region of interest, as can be observed from its poor \ac{mse} performance in Fig.~\ref{fig:zzb_vs_mmse_40}. In contrast to \ac{crb}, \ac{zzb} can reflect the ambiguity phenomenon. Consequently, the power spectral density of the \ac{zzb}-optimal waveform would have its mass more biased towards $B_{\rm dis}$ as the \ac{snr} increases, but will never become the impulse-like function corresponding to the \ac{crb}-optimal waveform, as portrayed in Fig.~\ref{fig:zzb_optimal_psd}. Therefore, despite that the \ac{zzb}-optimal waveforms do achieve their corresponding \acp{crb} at high \acp{snr}, the optimal \ac{crb} is not achievable for any finite \ac{snr}. As an ultimate limit, in the high-\ac{snr} regime, the minimum \ac{mse} is
\begin{equation}
\frac{1}{B}\Big(\int_0^B f^2B^{-1} {\rm d}f\Big)^{-1} = \frac{1}{3}
\end{equation}
that of the \ac{mse} achieved by the Sinc pulse.

We have seen that in the high-\ac{snr} regime, the \ac{rms} bandwidth is an important performance indicator, and provides a reasonable interpretation about the performance gain of the \ac{snr}-adaptive waveform over the Sinc pulse. However, similar arguments do not apply to the low-\ac{snr} regime. This can be seen from Fig.~\ref{fig:zzb_optimal_psd}, in which the \ac{rms} bandwidth of the waveform at ${\rm SNR_D}=10$dB is clearly smaller than that of the Sinc pulse, but it outperforms the Sinc pulse in the low-\ac{snr} regime ($\leq 15$dB), as can be observed from Fig.~\ref{fig:zzb_at_snrd} and Fig.~\ref{fig:zzb_vs_mmse_40}.

\begin{figure}[t]
\centering
\includegraphics[width=.45\textwidth]{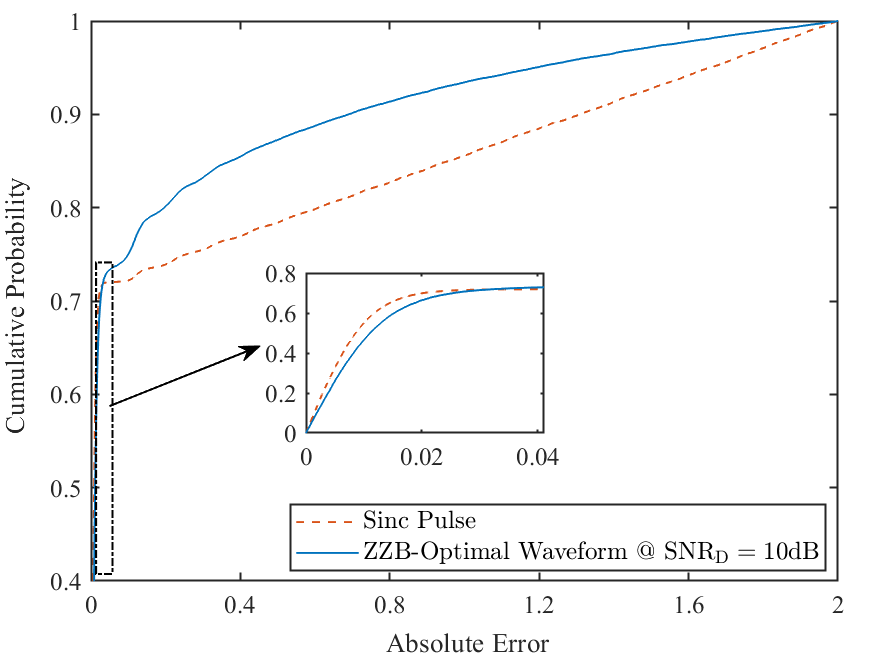}
\caption{The cumulative distribution of absolute ranging error at ${\rm SNR}=10$dB, obtained by employing the Sinc pulse and the \ac{zzb}-optimal waveform at ${\rm SNR}_{\rm D}=10$dB, respectively.}
\label{fig:abs_err_at_10}
\end{figure}

To further understand the low-\ac{snr} performance, in Fig.~\ref{fig:abs_err_at_10}, we plot the \ac{cdf} of the absolute ranging error at ${\rm SNR}=10$dB obtained by the \ac{zzb}-optimal waveform at ${\rm SNR_D}=10$dB, and that of the Sinc pulse, respectively. Observe that when the cumulative probability is larger than $0.7$, the \ac{cdf} of the Sinc pulse behaves similar to that of a uniform distribution, indicating that the Sinc pulse is essentially a ``random guess'' in this regime, clearly outperformed by the \ac{zzb}-optimal waveform. Interestingly, when the cumulative probability is less than $0.7$, the Sinc pulse outperforms the \ac{zzb}-optimal waveform (as seen in the zoomed-in figure). This phenomenon may be better illustrated by investigating the normalized \acp{acf} $\widetilde{R}(x)$ shown in Fig.~\ref{fig:zzb_optimal_acf}. It is seen that the \ac{acf} mainlobe of the \ac{zzb}-optimal waveform at ${\rm SNR}_{\rm D}=10$dB is wider than that of the Sinc pulse, which explains why the Sinc pulse outperforms the \ac{zzb}-optimal waveform when the cumulative probability is less than $0.7$ (corresponding to the case in which the distance estimation still fall in the mainlobe). Nevertheless, outside the mainlobe, the ${\rm SNR}_{\rm D}=10$dB waveform still has a considerable autocorrelation, much larger than that of the Sinc pulse. Consequently, the Sinc pulse would find the signal completely lost in a strong noise, whereas the ${\rm SNR}_{\rm D}=10$dB waveform still preserves some ranging information, since its relatively large \ac{acf} values beyond the mainlobe would enhance the signal detection probability. This reveals a design philosophy compared in stark contrast to that in the high-\ac{snr} regime:
\begin{remark}
In the low-\ac{snr} regime, instead of improving the resolution, enhancing the detection probability is of the highest priority. This suggests that appropriately widening the \ac{acf} can be beneficial.
\end{remark}

\section{Conclusions}
In this letter, we have proposed a ranging waveform design method capable of adpating to the \ac{snr}, based on the \ac{zzb}. The proposed design method is guaranteed to yield the globally \ac{zzb}-optimal waveform due to the convexity of the formulated problem.

Beyond the design method, the numerical results suggest that the waveform design philosophy in the low-\ac{snr} regime is strikingly different from that in the high-\ac{snr} regime: Essentially, the main performance indicator is the time-resolution in the high-\ac{snr} regime, whereas the detection probability becomes more crucial in the low-\ac{snr} regime. Therefore, larger \ac{rms} bandwidth would be favorable for higher \ac{snr}, whereas smaller \ac{rms} bandwidth can be beneficial when the \ac{snr} is relatively low, which improves the detection probability at the cost of sacrificing some resolution. Possible applications of the proposed method include pulse shaping for target ranging, and power allocation for OFDM-based ranging systems.

\appendices
\section{Proof of Proposition \ref{prop:convex_variational}}\label{sec:proof_convex_variational}
\begin{IEEEproof}
Note that for $x>0$, we have
\begin{equation}
Q(\sqrt{x}) = \frac{1}{2\sqrt{\pi}} \Gamma \Big(\frac{1}{2},\frac{1}{2}x\Big),
\end{equation}
where $\Gamma(s,t)$ denotes the incomplete Gamma function having the derivative of
$\frac{\partial \Gamma(s,t)}{\partial t} = -t^{s-1}e^{-t}$ \cite{table_integral}. Also observe that the expression of $\zeta_{\rm SNR_D}[\widetilde{R}]$ does not contain the derivatives of $\widetilde{R}(x)$ with respect to $x$. Therefore, according to the Euler-Lagrange equation, the first-order variation of $\zeta_{\rm SNR_D}[\widetilde{R}]$ with respect to $\widetilde{R}$ is given by
\begin{align}
\delta \zeta_{\rm SNR_D} &= \int_0^{\epsilon_{\max}} \left[x\frac{{\rm d}}{{\rm d}\widetilde{R}}Q\left(\sqrt{2^{-1}{\rm SNR}_{\rm D}(1-\widetilde{R})}\right)\right]\delta\widetilde{R} {\rm d} x \nonumber\\
&=\frac{x{\rm SNR}_{\rm D}}{8\sqrt{\pi}} \int_0^{\epsilon_{\max}} \frac{e^{-\frac{\rm SNR_D}{4}(1-\widetilde{R})}}{\sqrt{\frac{\rm SNR_D}{2}(1-\widetilde{R})}}\cdot \delta \widetilde{R}{\rm d}x.
\end{align}
The second-order variation takes the following form
\begin{align}\label{second_variation}
\delta^2\zeta_{\rm SNR_D}&=\frac{x\sqrt{\rm SNR_D}}{4\sqrt{2\pi}}\int_0^{\epsilon_{\max}} e^{-\frac{\rm SNR_D}{4}(1-\widetilde{R})} (\delta\widetilde{R})^2\nonumber \\
&\hspace{3mm}\times \left(\frac{\rm SNR_D}{4}(1-\widetilde{R})^{\frac{1}{2}} +\frac{1}{2}(1-\widetilde{R})^{-\frac{3}{2}}\right) {\rm d}x.
\end{align}
For $x\neq 0$, we have $1-\widetilde{R}(x)>0$. Thus for $x\in (0,\epsilon_{\max}]$, the second-order variation satisfies $\delta^2\zeta_{\rm SNR_D}>0$ for any $\delta\widetilde{R}$, implying that the functional $\zeta_{\rm SNR_D}[\widetilde{R}]$ is convex.
\end{IEEEproof}

\bibliographystyle{IEEEtran}
\bibliography{IEEEabrv,zzb}

\end{document}